\documentclass[preprint]{aastex61}
\usepackage{epstopdf}
\usepackage{commath}
\usepackage{color}
\usepackage{hyperref}
\usepackage{mathtools}
\usepackage[toc,page]{appendix}
\usepackage{bm}

\graphicspath{{./}{figures/}}

\received{\today}

\submitjournal{ApJ}
\shorttitle{}
\shortauthors{C.A. González et al. 2023}
\begin{document}

\title{Local proton heating at  magnetic discontinuities in Alfv\'enic and non-Alfv\'enic solar wind}  
\correspondingauthor{C.A. González}
\email{carlos.gonzalez1@austin.utexas.edu}

\author{C.A. González}
\affiliation{Department of Physics, The University of Texas at Austin, Austin, TX USA}

\author{J. L. Verniero}
\affiliation{Code 672, NASA, Goddard Space Flight Center, Greenbelt, MD 20771, USA}

\author{R. Bandyopadhyay}
\affiliation{Department of Astrophysical Sciences, Princeton, New Jersey 08544, USA}

\author{A. Tenerani}
\affiliation{Department of Physics, The University of Texas at Austin, Austin, TX USA}


\begin{abstract}
 We investigate the local proton energization at  magnetic discontinuities/intermittent structures and the corresponding kinetic signatures in velocity phase space in Alfvénic (high cross helicity) and non-Alfvénic (low cross helicity) wind streams observed by Parker Solar Probe. By means of the Partial Variance of Increments method, we find that the hottest proton populations are localized around compressible, kinetic-scale magnetic structures in both types of wind. Furthermore, the Alfvénic wind shows preferential enhancements of $T_\parallel$ as smaller scale structures are considered, whereas the non-Alfvenic wind shows preferential $T_\bot$ enhancements. Although proton beams are present in both types of wind, the proton velocity distribution function displays distinct features. Hot beams, i.e., beams with beam-to-core perpendicular temperature $T_{\bot,b}/T_{\bot,c}$ up to three times larger than the total distribution anisotropy, are found in the non-Alfvénic wind, whereas colder beams in the Alfv\'enic wind. Our data analysis is complemented by 2.5D hybrid simulations in different geometrical setups, which support the idea that proton beams in Alfv\'enic and non-Alfv\'enic wind have different kinetic properties and different origins. The development of a perpendicular nonlinear cascade, favored in balanced turbulence, allows a preferential relative enhancement of the perpendicular plasma temperature and the formation of hot beams. Cold field-aligned beams are instead favored by Alfv\'en wave steepening. Non-Maxwellian  distribution functions are found near discontinuities and intermittent structures, pointing to the fact that the nonlinear formation of small scale structures is intrinsically related to the development of highly non-thermal features in collisonless plasmas.  
 Our results contribute to understanding the role of different coherent structures in proton energization and their implication in collisionless energy dissipation processes in space plasmas.

\end{abstract}


\section{Introduction}
 
Magnetic discontinuities have been observed for a long time throughout the heliosphere~\citep{colburn1966,parker1994spontaneous,Burlaga91,TsurutaniEA99,
VasquezEA2007},  and particularly in turbulent solar wind streams where discontinuities have been related to intermittent structures \citep{greco2016complex,greco2008}. Magnetic discontinuities and intermittent structures may play an important role in energy dissipation and transport, and in particle acceleration~\citep{Osman_2010,OsmanEA2012,Tessein_2013}. Several mechanisms have been proposed for solar wind heating, including resonant wave-particle interactions (such as Landau damping and ion-cyclotron resonance \citep{narita2015kinetic,gary2008damping,sahraoui2010three,bowen2022situ}), magnetic pumping \citep{lichko2017magnetic},  stochastic heating \citep{chen2001resonant,johnson2001stochastic,ChandranEA2010,vech2017nature} and intermittent dissipation \citep{dmitruk2004test,karimabadi2013coherent,osman2014magnetic}. However, solar wind streams differ by the type of fluctuations and turbulence, and which heating mechanisms are favored under different solar wind conditions remains to be understood. 

Tangential (TD) and rotational discontinuities (RD) are the most common types of discontinuities observed in the solar wind \citep{Neugebauer2006,Paschmann2013}. In theory, TDs and RDs can be identified depending on the normal component of the magnetic field and the changes in plasma parameters across the structure \citep{hudson1970discontinuities}. TDs are stationary structures in the plasma frame and have zero normal magnetic field component. These structures represent boundaries between different plasma parcels and are in pressure balance, without restrictions on the variations of magnetic field magnitude across the discontinuity. TDs are commonly found at reconnection sites or, more generally, at the boundaries of flux tubes \citep{Greco_2009,ServidioEA2011,Zhdankin_2012}. On the other hand, RDs have a non-zero normal magnetic field component and a field-aligned flow. RDs are propagating structures that have been associated with phase-steepened Alfvén waves~\citep{tsurutani1994,MedvedevEA1997,VasquezEA01}. 

The classification of TDs and RDs is not straightforward, not only due to single-spacecraft limitations to estimate accurately the normal component, but also because discontinuities are non-ideal and can display properties of both TDs and RDs~\citep{neugebauer1984,horbury2001three,knetter2004four,artemyev2019kinetic}. Nevertheless, previous work has shown that different types of magnetic structures are statistically detected in different types of wind. Namely, compressible structures are most likely observed in the slow solar wind, while structures with small compressibility are typically observed in the fast solar wind~\citep{perrone2016compressive,perrone2017coherent}.  Although non-Alfv\'enic and Alfv\'enic solar wind is traditionally associated with slow and fast streams, recent observations showed the existence of “slow Alfvenic” wind \citep{ DAmicis2021}. This suggests that the solar wind cannot be simply categorized based solely on its velocity. The normalized residual energy (the relative energy in kinetic and magnetic fluctuations), the Alfv\'en ratio (the ratio between kinetic to magnetic fluctuations), and the normalized cross helicity (the correlation between magnetic and velocity field fluctuations) are some of the measures that quantify Alfvénicity. In this work, we classify the solar wind based on the average value of cross-helicity. We consider the solar wind as Alfvénic when it contains fluctuations with cross helicity close to unity. In contrast, the non-Alfvénic wind refers to fluctuations with cross helicity approaching zero.  

Several studies have shown that plasma temperature enhancements are associated with intermittent structures/discontinuities identified via the partial variance of increment (PVI) method \citep{Osman_2010,Qudsi_2020,sioulas2022,phillips2023}. The PVI method enables identification of current sheet structures, but it does not distinguish whether those structures are predominantly compressible or rotational. However, TDs and RDs are fundamentally different types of structures that affect particle energization in different ways, and how compressible (TD-like) and incompressible (RD-like) structures in the solar wind interact with particles ultimately leading to heating (and their velocity-space signatures) remains to be investigated. 

In this work, we address such a problem by investigating proton heating at different types of intermittent structures and/or discontinuities in the inner heliosphere, by using Parker Solar Probe (PSP) data and 2.5D hybrid-kinetic simulations. Toward this goal, we conduct a comparison study between Alfvénic and non-Alfvénic streams and investigate the dependence of temperature anisotropies on Alfv\'enicity and on the type of magnetic structures. Additionally, hybrid simulations, with simplified geometry settings, are used to investigate what processes are likely to contribute to proton heating and to the generation of the observed proton temperature anisotropy in each type of~wind.  

This paper is organized as follows. In section \ref{sect2}, we describe the data and methods, and we report our data analysis. An overview of the properties of magnetic field fluctuations in Alfv\'enic and non-Alfv\'enic wind is reported in  section~\ref{overview_data}. In section \ref{PVI_data} we discuss the correlation between PVI and proton temperatures and temperature anisotropy, and in sec.~\ref{proton_kinetic_PSP} we consider two case studies to show the typical signatures in proton velocity space at different magnetic structures in the two types of wind. Results from numerical simulations and a comparison between numerical outputs and PSP data are reported in section \ref{sect3}. The summary and discussion are in section \ref{sect4}.

\section{PSP observations}
\label{sect2}

\subsection{Overview of fluctuations' properties for E6-E9}
\label{overview_data}

In this work we have used PSP data from Encounter 6 through Encounter 9 (E6-E9), which occurred from September 9, 2020, to August 15, 2021 covering a range of radial distances from the sun  $0.07< R< 0.263$~AU. We use magnetic field data from the flux-gate magnetometer at 4 samples/cycle resolution (FIELDS; \citet{bale2016fields}). The 3D proton velocity distribution functions (L2) and its moments (L3) at 3s resolution are obtained by the electrostatic analyzer (SPAN-I; \citet{Livi_2022}), part of the Solar Wind Electrons Alphas and Protons instrument suite (SWEAP; \citet{kasper2016solar}); Since SPAN-I is partially obstructed by PSP's thermal protection shield, we use the quasi-thermal noise (QTN) measured by the FIELDS Radio Frequency Spectrometer (\cite{moncuquet2020}) to have a more accurate determination of plasma density.  The total parallel and perpendicular proton temperatures, $T_\bot$ and $T_\parallel$, are obtained by projecting the temperature tensor in the parallel and perpendicular directions with respect to the magnetic field.

\begin{figure}[h!]
  \centering
  \includegraphics[width=0.98\textwidth]{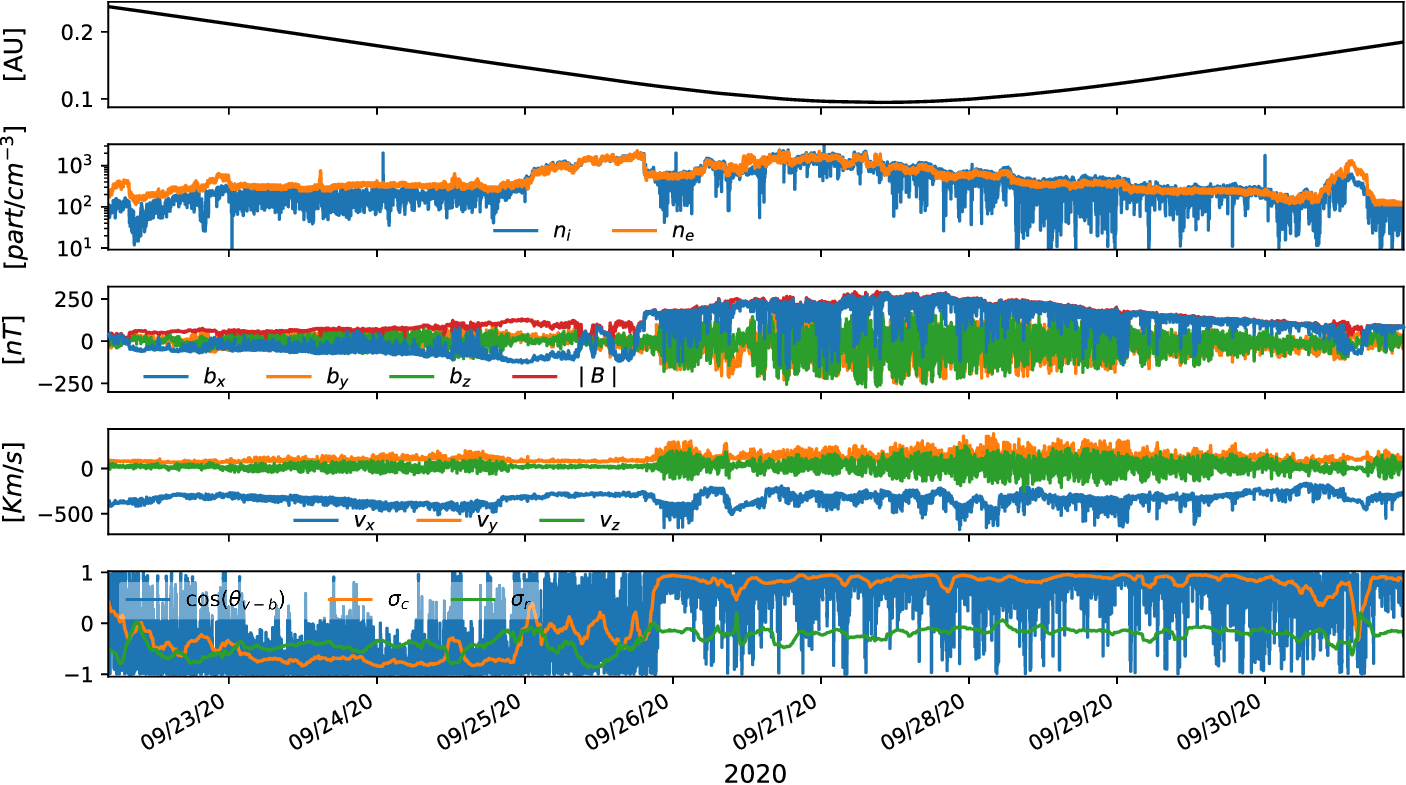}
  \caption{PSP Encounter 6 data: spacecraft radial distance $R$ (top panel); proton and electron number density $n_{i,e}$ (second panel); magnetic field ${\bf b}$, its magnitude $B$ and proton bulk velocity ${\bf V}$ in the instrument frame (third and fourth panels); normalized cross-helicity $\sigma_c$, residual energy $\sigma_r$ and the cosine of the angle between the velocity and magnetic field fluctuations $\cos\theta_{vb}$ (bottom panel).}
\label{fig:one}
\end{figure}

For reference, in Fig.~\ref{fig:one} we provide an overview of E6. The top four panels show the spacecraft radial distance ($R$), the proton and electron number density ($n_{i,e}$), and the magnetic field (${\bf b}$ and $B=|{\bf b}|$) and proton bulk velocity (${\bf V}$) in the instrument frame ($x$-component pointing toward the sun). The fifth panel shows the Alfvénic properties of the wind represented by the normalized cross-helicity $\sigma_c$ and residual energy $\sigma_r$ of fluctuations, given respectively by

\begin{equation}
 \sigma_c = \frac{ \langle | {\bf z}^{+} |^2 \rangle - \langle | {\bf z}^{-} |^2  \rangle}{\langle | {\bf z}^{+} |^2  \rangle  + \langle | {\bf z}^{-} |^2  \rangle  }
   \quad\text{and}\quad 
\sigma_r =  \frac{\langle | \mathbf{\delta V} |^2 \rangle - \langle | \mathbf{\delta b} |^2\rangle/(\mu_0 \rho)} { \langle | \mathbf{\delta V} |^2\rangle + \langle | \mathbf{\delta b} | ^2\rangle/(\mu_0\rho)},
\end{equation}

where brackets $\langle ... \rangle$ denote the moving average over a time window,  ${\bf z}^{\pm} = \mathbf{\delta V} \pm \mathbf{\delta b}/\sqrt{\mu_0 \rho}$ are the Els{\"a}sser variables defined by the velocity and magnetic field fluctuations with respect to the mean field and by the  QTN mass density $\rho$ assuming quasi-neutrality. In this work we fix the averaging window to 2 hours, that corresponds to several times the correlation time of the magnetic field in the inner heliosphere \citep{Parashar_2020,sioulas2022}. The cross-helicity measures the relative dominance of inward and outward propagating Alfv\'enic fluctuations (${\bf z}^{\pm}$), and the residual energy quantifies the partition between the fluctuations' kinetic and magnetic energy. The bottom panel of Fig.~\ref{fig:one} shows the cosine of the angle $\theta_{vb}$ between the velocity and magnetic field, defined as 
\begin{equation}
\cos{\theta_{vb}} = \frac{\mathbf{\delta V} \cdot \mathbf{\delta b}}{|\mathbf{\delta V}| |\mathbf{\delta b}|}. 
\end{equation}

As will be discussed below,  fluctuations during E6 are remarkably Alfv\'enic, with high cross-helicity intervals (defined by $|\sigma_c| > 0.75$) that last several days (particularly during and after perihelion). In addition, a shorter interval of low-Alfvénicity is observed in the vicinity of the heliospheric current sheet (September 25th and 26th, 2020), a complex structure with dominant magnetic energy and nearly zero cross-helicity. This agrees with the observed non-Alfv\'enic wind reported near the Heliospheric Current Sheet (HCS) in E4 \citep{chen2021near}.

\begin{figure}
  \centering
  \includegraphics[width=\textwidth]{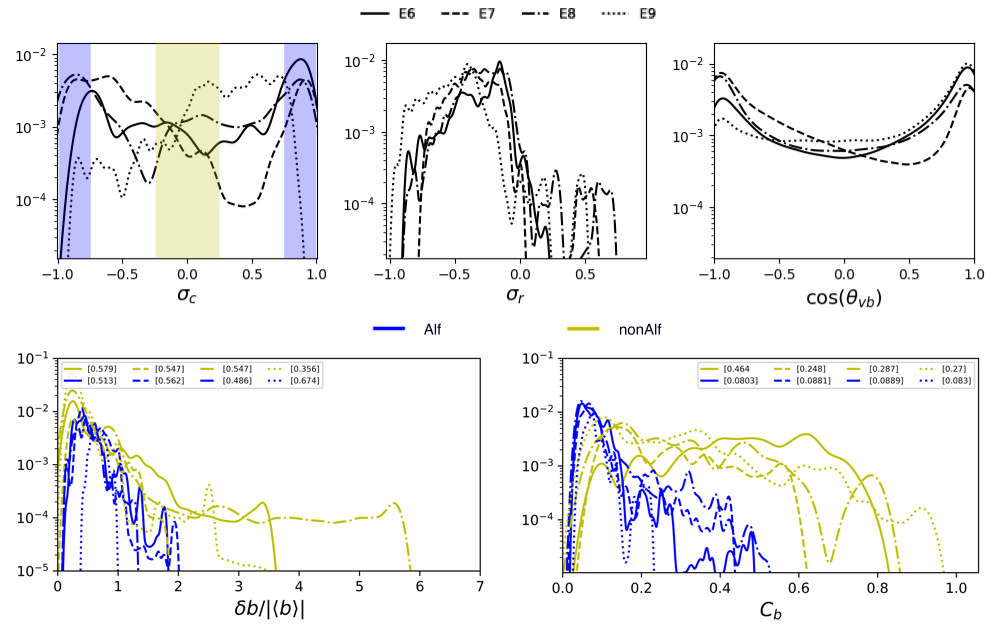}
  \caption{Fluctuations' properties for E6-E9. Top panels: PDF of normalized cross-helicity $\sigma_c$(top-left panel), residual energy $\sigma_r$ (middle panel) and the cosine angle of velocity and magnetic field fluctuations $\cos\theta_{vb}$ (top-right panel). Different Encounters correspond to a different line style as indicated in the legend at the top. The shaded blue and yellow areas indicate values $|\sigma_c|>0.75$ and $|\sigma_c|<0.25$, respectively. Bottom panels: PDF of the normalized fluctuations amplitude $\delta b/|\langle {\bf b}\rangle|$ (bottom left) and of the ratio of magnetic field and magnetic field strength variances $C_b$ (bottom right) for the Alfv\'enic wind (blue lines, corresponding to the blue shaded area in the top left panel) and for the non-Alfv\'enic wind (yellow lines, corresponding to the yellow shaded area). The values in the legend of the bottom panels shows the mean value of each distribution.}
\label{fig:two}
\end{figure}

In Fig.~\ref{fig:two} we show the statistical analysis of Alfvénic properties and compressibility of solar wind fluctuations from E6 through E9. The top panels show the Probability Density Function (PDF) of $\sigma_c,\sigma_r$ and $\cos{\theta_{ub}}$ for each Encounter. Although data display a range of values $-1 \leq \sigma_c \leq 1$, intervals of imbalanced fluctuations with $|\sigma_c|>0.5$ can be identified, with peaks of the distribution of $\sigma_c$ at $|\sigma_c|\geq0.75$  (see also Fig.~\ref{fig:one}). In general, fluctuations are magnetically dominated, a trend which is consistent throughout all Encounters \citep{chen2020evolution,shi2021alfvenic}, with mean  value $\sigma_r\approx
 -0.5$ and with only a few intervals being characterized by an excess of kinetic energy.  Lastly, the PDF of $\cos{\theta_{ub}}$ peaks at $\pm 1$, corresponding to fluctuations with aligned ${\bf V}$ and ${\bf B}$ in all the Encounters. The Alfv\'enic wind (those intervals with $|\sigma_c|\geq0.75$) present strong alignment, i.e., $| \cos \theta_{v b} | \sim 0.89$, while the non-Alfv\'enic wind (those intervals with $|\sigma_c|\leq 0.25$) presents a moderate alignment with a broader distribution of angles having mean value $| \cos \theta_{v b} | \sim 0.57$. In the latter case, such alignment corresponds with the relaxation and suppression of nonlinearities in standard balanced turbulence that has been previously observed in both simulations and solar wind measurement \citep{matthaeus2008rapid,matthaeus2012review}.

The bottom panels of Fig.~\ref{fig:two} show the PDF of the amplitude of  magnetic field fluctuations, 
\begin{equation}
    \frac{\delta b}{|\langle{\bf b}\rangle|} \coloneqq \sqrt{\frac{\langle(b_x -\langle b_x \rangle)^2 + (b_y -\langle b_y \rangle)^2 + (b_z -\langle b_z \rangle)^2 \rangle}{\langle b_x\rangle^2+\langle b_y\rangle^2+\langle b_z\rangle^2}},
\end{equation} 

and of the ratio of the variances of the magnetic field and magnetic field strength \citep{villante1982radial},  
\begin{equation}
   C_b\coloneqq \sqrt{\frac{\langle (B-\langle B \rangle)^2\rangle}{\langle(b_x -\langle b_x \rangle)^2 + (b_y -\langle b_y \rangle)^2 + (b_z -\langle b_z \rangle)^2\rangle}},
\end{equation}

the latter providing a proxy for magnetic compressibility. In this analysis, we have sorted the solar wind into Alfvénic (intervals with $|\sigma_c|  > 0.75$, corresponding to the blue-shaded areas in Fig.~\ref{fig:two}, top left) and non-Alfv\'enic wind (intervals with $ |\sigma_c|  < 0.25$, corresponding to the yellow-shaded area). For the non-Alfvénic wind we have also removed data corresponding to the HCS crossing, where $|\langle {\bf b}\rangle|\approx0$ (for the time window considered of 2-hours), to avoid fictitious long tails of the PDF.

From Fig.~\ref{fig:two}, bottom left panel, it can be observed that the distributions of the normalized fluctuations' amplitude are remarkably different between Alfvénic (blue) and non Alfvénic (yellow) wind. In the former case, relative fluctuations' amplitudes are bounded roughly by $\delta b/|\langle{\bf b}\rangle|\lesssim 2$. This is in line with previous observations of Alfvénic fluctuations in Helios and Ulysses, where saturation of amplitudes (although defined differently) was found, a feature that is consistent with spherically polarized fluctuations \citep{matteini20181}. In the non-Alfvénic case, instead, relative fluctuations' amplitudes are large and there is no constraint on their value (E9 being the only ambiguous case). We interpret the different PDF of the normalized amplitude as a signature of a higher level of compressibility in the non-Alfvénic wind (at the timescale of 2 hours). The bottom right panel of Fig.~\ref{fig:two} shows the distribution of $C_b$ for both types of wind.  As expected, $C_b$ in the non-Alfvénic wind is systematically larger than in the Alfvénic wind, with the PDF mean in the non-Alfvénic wind reaching up to 5 times that of the Alfv\'enic wind.

To summarize,  we have considered the statistical properties of fluctuations over a timescale of 2 hours. From this part of our data analysis, we conclude that fluctuations in the Alfvénic ($|\sigma_c|>0.75$) and non-Alfvénic ($|\sigma_c|<0.25$) wind at distances $R<0.25$~AU display statistical properties similar to those traditionally observed further away from the sun, namely, almost incompressible and saturated fluctuations in the Alfvénic wind and highly compressible fluctuations in the non-Alfvénic wind.

\subsection{Correlations between PVI and proton thermal properties in different types of wind}
\label{PVI_data}

To study the connection between proton heating and discontinuities or, more generally, intermittent structures in different plasma conditions, we utilized the Partial Variance of Increments (PVI) method by calculating the following quantity~\citep{greco2008intermittent}:
\begin{equation}
\label{eq:PVI}
PVI = \frac{| \Delta \mathbf{B}(t,\tau) |}{\sqrt{\langle | \Delta \mathbf{B}(t,\tau) |^2 \rangle }},
\end{equation}

where $ \Delta \mathbf{B}(t,\tau) = \ \mathbf{b}(t+\tau) - \mathbf{b}(t)$ is the magnetic field increment vector. The PVI method identifies non-Gaussian features in the magnetic field and events with values $PVI>3$ are typically associated with coherent structures or discontinuities. Although the PVI defined in Eq.~\ref{eq:PVI} captures variations in the magnetic field, it does not distinguish RDs, TDs or switchbacks, that sometimes display properties of both RDs and TDs~\citep{larosa2021switchbacks}. For this reason, we also considered the PVI of the magnetic field strength, hereafter called $MAG$-$PVI$, by taking the variation of $B$, i.e., by substituting $\Delta {\bf B}(t,\tau)\Rightarrow\Delta B(t,\tau)=B(t+\tau) -B (t)$ in Eq.~(\ref{eq:PVI}). A comparison of events identified with $PVI$ and $MAG$-$PVI$ allows us to gain insights on whether compressible or incompressible magnetic structures are statistically associated to local particle energization. Finally, because of the resolution of SPAN-i data,  we used a time lag of $\tau = 12 \ s$, that lies within the range of scales where the transition from the inertial to the kinetic regime takes place, to correlate discontinuities at kinetic scales and proton heating (or variations of plasma moments). 

{After constructing the time series of $PVI$ (and $MAG$-$PVI$), we computed the normalized probability distribution PDF of various functions of temperature $f(T)$ conditioned on a given range of $PVI$ (and $MAG$-$PVI$) values 
\begin{equation}
    \mathcal{PDF}(f(T)|\theta_1\leq PVI\leq\theta_2).
\end{equation}

We have calculated the conditional PDF of the $(i)$ total proton temperature ($f(T)=T$), $(ii)$  variation of temperature over the time lag  $\tau=12\, s$ ($f(T)= T(t+\tau)-T(t)\equiv \delta T$), $(iii)$ parallel and perpendicular temperature ($f(T)=T_\parallel,\,T_\perp$ where $\parallel$ and $\bot$ are defined with respect to the local magnetic field direction $\mathbf{\hat {b}}=\mathbf{b}/B$), and $(iv)$ temperature anisotropy ($f(T)=T_\perp/T_\parallel$). We performed the same PVI analysis with  $\tau=1 \ s$ and then  resampled the PVI data to match the SPAN-i resolution, but we did not find significant quantitative differences.

\begin{figure}
  \centering
  \includegraphics[width=\textwidth]{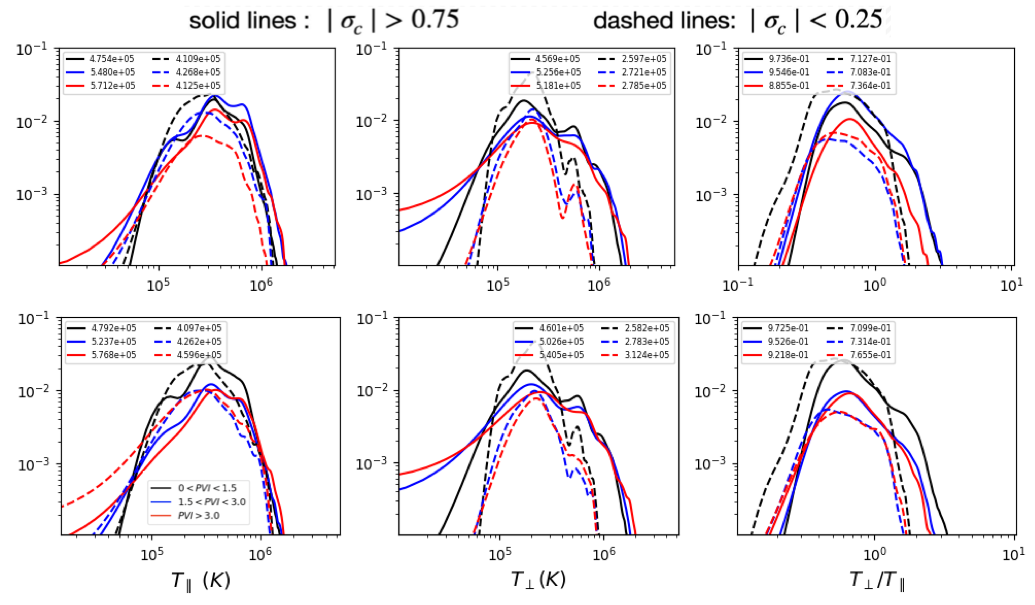}
  \caption {Conditional PDF with respect to $PVI$ (top panels) and $MAG$-$PVI$ (bottom panels) of proton temperatures using data from E6 through E9: PDF of parallel proton temperature $T_\parallel$ (left panels); PDF of perpendicular proton temperature $T_\bot$ (middle panels); PDF of proton temperature anisotropy $T_\bot/T_\parallel$ (right panels). Solid lines correspond to Alfv\'enic wind and dashed lines to non-Alv\'enic wind. Different color lines show  PDF that corresponds to different ranges of $PVI$ ($MAG$-$PVI$): black lines ($0<PVI\le1.5$), blue lines ($1.5<PVI\le3.0$), and red lines ($PVI>3.0$). The numbers on the legends indicate the mean value of each distribution.}
\label{fig:three}
\end{figure}  

In agreement with other studies, we find that the highest values of $T$ and $\delta T$ are associated with the highest PVI values in both winds (not shown here; see also, e.g., \cite{Osman_2010,Qudsi_2020,sioulas2022}), supporting the idea that heating occurs at localized structures. Results of our PVI analysis are reported in Fig.~\ref{fig:three} for parallel and perpendicular temperatures. 

Fig.~\ref{fig:three} shows the conditional PDF with respect to $PVI$ (top panels) and $MAG$-$PVI$ (bottom panels) of $T_\parallel$ (left), $T_\bot$ (middle) and $T_\bot/T_\parallel$ (right). The PDFs have been obtained by combining data from all Encounters (E6-E9) and by using the cross helicity to sort Alfv\'enic  ($|\sigma_c|>0.75$, solid lines) and non-Alfvénic ($|\sigma_c|<0.25$, dashed lines) intervals. Red, blue and black colors correspond to different ranges of $PVI$ ($MAG$-$PVI$), that we split into  $(0 < PVI \leq 1.5)$, $(1.5 < PVI \leq 3.0)$, and $( PVI > 3.0)$ respectively. The legend in each panel reports the mean value of each distribution. Finally, E6 was also analyzed individually, but we found the same trends as those shown in Fig.~\ref{fig:three} and discussed below.

As can be seen from the PDFs of $T_\bot/T_\parallel$ (Fig.~\ref{fig:three}, right panels), the non-Alfvénic wind is more anisotropic than Alfvénic wind and, on average,  $(T_\bot/T_\parallel)_{non Alfv} < (T_\bot/T_\parallel)_{Alfv} < 1$. As we will see, such an anisotropy is due to the ubiquitous presence of proton beams. While proton beams have been commonly observed at all radial distances in the Alfv\'enic wind before PSP (e.g., ~\cite{marsch2012helios}), the frequent observation of field-aligned beams in the non-Alfv\'enic wind (including the so-called ``hammerhead'' distributions; \cite{Verniero_2020, verniero2022strong}) is one of the unexpected results of PSP. Further inspection of the temperature anisotropy PDFs shows that Alfv\'enic and non-Alfv\'enic wind display an opposite correlation between $PVI$ (and $MAG$-$PVI$) values, and $T_\bot/T_\parallel$. For the $PVI$ case (Fig.~\ref{fig:three}, top right panel),  the mean values of the non-Alfv\'enic wind anisotropy lie in the range $0.71\lesssim T_\perp/T_\parallel\lesssim 0.73$ with increasing $PVI$ (dashed lines). In the Alfv\'enic wind (solid lines), we find instead $0.88 \lesssim T_\perp/T_\parallel\lesssim0.97$ with decreasing $PVI$. The $MAG$-$PVI$ case (Fig.~\ref{fig:three}, bottom right panel) shows similar values of anisotropy and trends.

Insights on parallel and perpendicular temperature can be found by cross-checking PDFs of $T_\bot/T_\parallel$ with the PDFs of $T_\parallel$ and $T_\bot$ (Fig.~\ref{fig:three}, left and middle panels). Interestingly, we find that the ensembles of events selected through each $MAG$-$PVI$ value have consistently higher average temperatures than the ensembles of events selected with the traditional $PVI$ values. With increasing $MAG$-$PVI$ values, the average temperatures are in the range $(4.792<T_\parallel<5.768)\times 10^5$~K and $(4.601<T_\bot<5.405)\times 10^5$~K in Alfv\'enic wind. In the non-Alfv\'enic wind we find $(4.097<T_\parallel<4.596)\times 10^5$~K and $(2.582<T_\bot<3.124)\times 10^5$. By contrast, the largest temperatures found with $PVI>3$ are $T_\parallel=5.358\times10^5$~K and $T_\bot=4.569\times10^5$~K (Alfv\'enic), and $T_\parallel=4.125\times10^5$~K and $T_\bot=2.785\times10^5$~K (non Alfv\'enic).  

To quantify the trends of perpendicular and parallel temperature with $MAG$-$PVI$, we calculated the relative temperature enhancement between the smallest and the largest range of $MAG$-$PVI$ considered. In the Alfv\'enic wind we find $\Delta T_\parallel/T_\parallel\simeq 20\%$ and $\Delta T_\bot/T_\bot\simeq 17\%$, indicating that the relative enhancements of parallel temperature are larger than those in the perpendicular temperature as smaller scales are considered. The opposite trend is found in the non-Alfv\'enic wind, where $\Delta T_\parallel/T_\parallel\simeq 12\%$ and $\Delta T_\bot/T_\bot\simeq 21\%$. Such an opposite trend of temperature enhancements explains why the anisotropy $T_\bot/T_\parallel$ shifts away from (towards) unity as PVI values increase for the Alfv\'enic (non-Alfv\'enic) wind.

From this analysis we conclude, first, that structures with the largest variations in $B$ (magnetic compressible structures) contribute to the highest $T_\parallel$ and $T_\bot$ in both types of wind, suggesting that the largest temperature enhancements occur in compressible structures also in the Alfv\'enic wind. Second, the Alfv\'enic wind shows a preferential enhancement of $T_\parallel$ as smaller scale structures are considered, whereas the non-Alfv\'enic wind shows a preferential enhancement of $T_\bot$.  

\subsection{Proton kinetic features at different magnetic structures}
\label{proton_kinetic_PSP}

\begin{figure}
  \centering
  \includegraphics[width=0.95\textwidth]{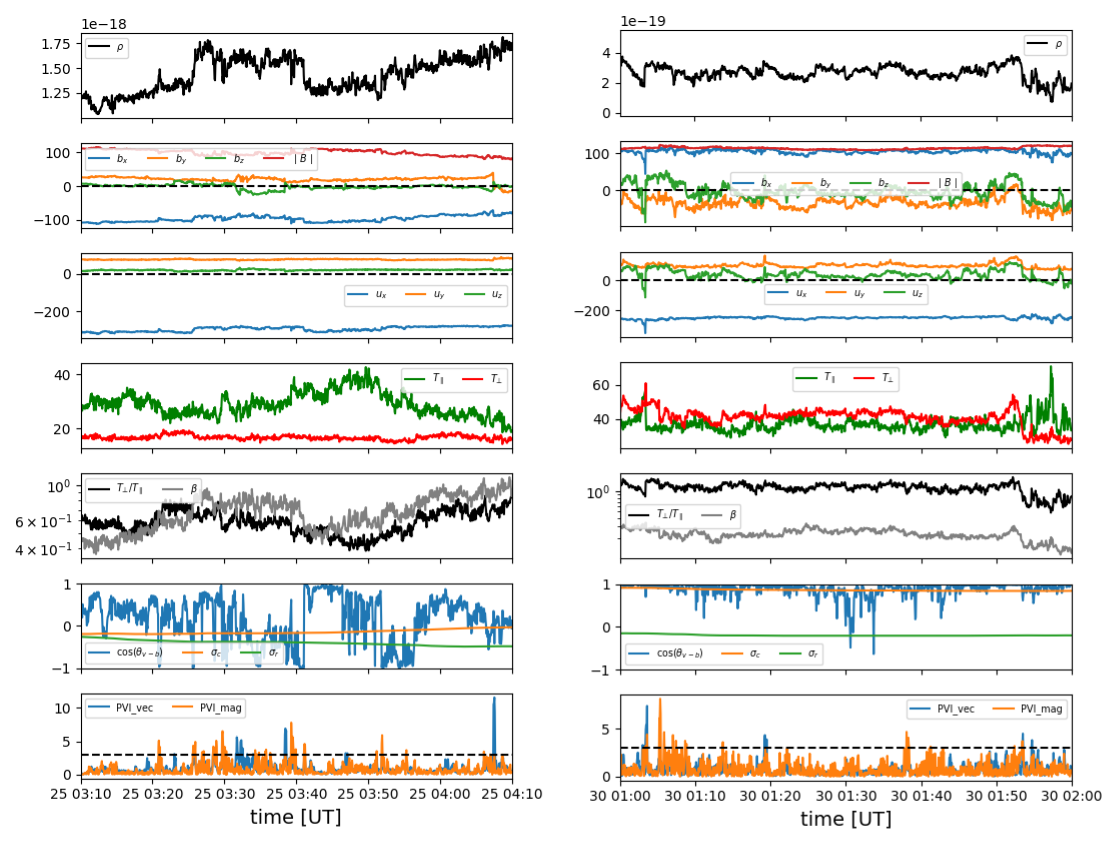}
  \caption{Data from two intervals during E6. The non-Alfv\'enic interval is shown on the left panels and the Alfvénic one on the right. From top to bottom: proton mass density ($\rho \ \  [ kg/m^{3}]$); magnetic field (${\bf b}$ and $B \ [nT] $);  proton bulk velocity (${\bf V} \  \ [km/s^{-1}]$); parallel and perpendicular temperature ($T_\parallel$ and $T_\bot \ \   [eV]$); temperature anisotropy ($T_\bot/T_\parallel$) and plasma beta ($\beta$). The last two panels show fluctuations properties ($\sigma_r$, $\sigma_c$ and $\cos\theta_{vb}$) and the PVI values.}
\label{fig:five}
\end{figure}

To investigate the local energization of protons at discontinuities/intermittent structures and kinetic features of the distribution function in velocity space, we manually inspected 1-hour intervals during quiet solar wind and without switchbacks. We selected two sub-intervals that characterize the Alfvénic and the non-Alfv\'enic wind (according to high and low cross-helicity), respectively. The two selected sub-intervals occurred during E6 at approximately the same radial distance from the sun ($30$ solar radii). Different fields and particle quantities are reported in Fig. \ref{fig:five}, where panels on the left correspond to the non-Alfvénic wind and those on the right to the Alfvénic wind.

The non-Alfv\'enic sub-interval occurred near the heliospheric current sheet crossing, on September 25th, 2020 from 03:10:00 to 04:10:00 UT, with an average cross-helicity $ \sigma_c \simeq -0.25$. This interval presents many discontinuities with large PVI values. The structures are substantially compressible with variations of $B$ and  $\beta$ enhancements.  

\begin{figure}
  \centering
  \includegraphics[width=0.95\textwidth]{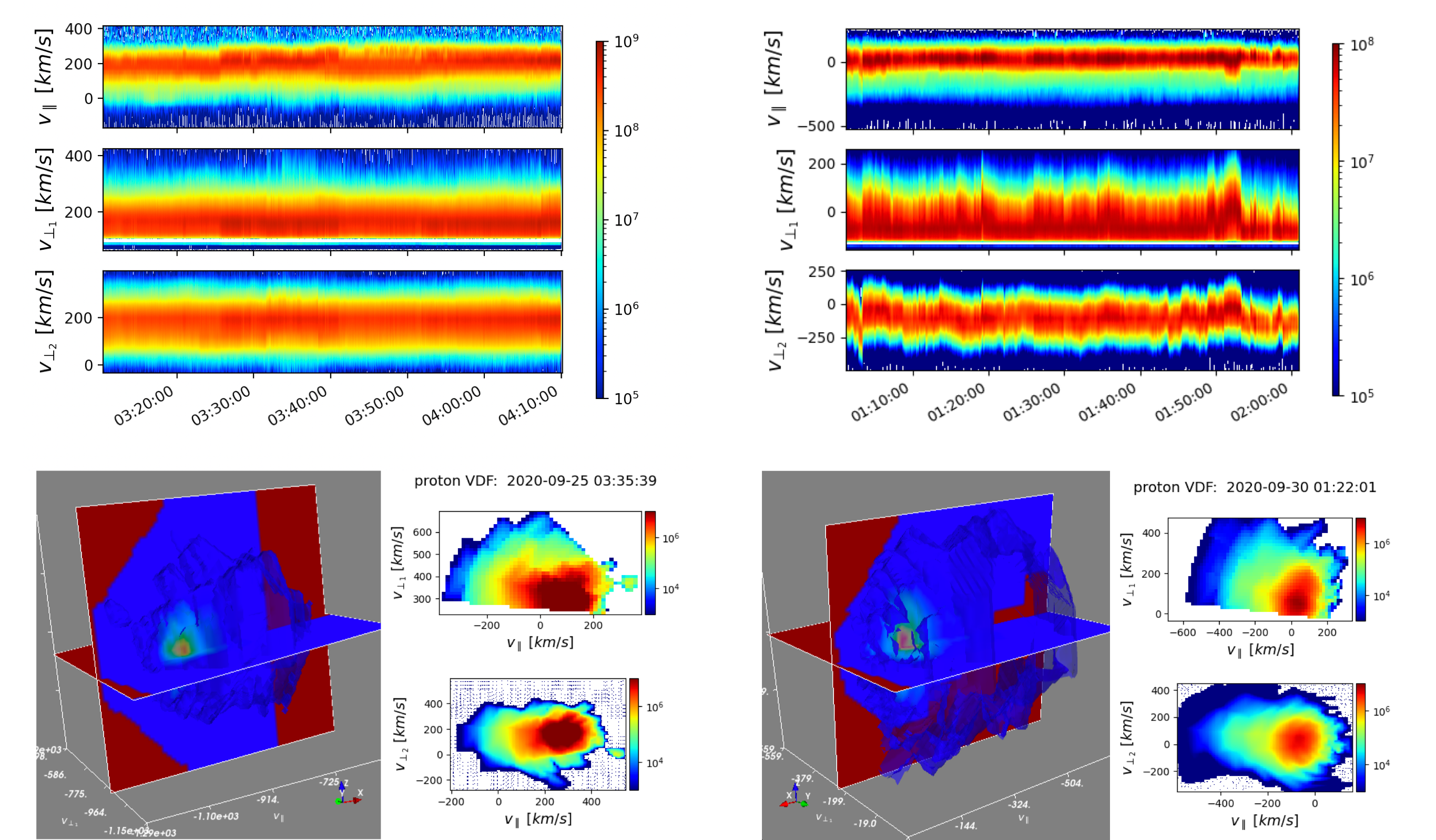}
  \caption{The reduced proton VDF as a function of parallel ($f(t,v_\parallel)$) and perpendicular ($f(t,v_{\bot_{1,2}})$) velocity components for the non-Alfv\'enic  (left column) and Alfv\'enic (right column) sub-intervals. The bottom panels show two examples of the reconstructed 3D VDF $f(v_\parallel, v_{\bot_{1}},v_{\bot,2})$ and its 2D projections, $f(v_\parallel, v_{\bot_{1,2}})$, at the time of an event with $PVI>3$ in each type of wind.} 
\label{fig:six}
\end{figure} 

The Alfvénic interval occurred on September 30th, 2020 between 01:00:00 to 02:00:00 UT with an overall $\abs{\sigma_c}\simeq 0.88$. The fluctuations show the typical highly-correlated velocity-magnetic field that characterizes wave-like fluctuations with rotation of $\cos \theta_{ub}$ near small-scale structures. In general, the variations of the magnitude of the magnetic field and the PVI values are smaller compared to the discontinuities in the non-Alfv\'enic wind. In both cases, however, the most substantial variations in proton temperature occur within large $PVI$ and $MAG$-$PVI$ structures and correspond to a net enhancement of $T_\parallel/T_\bot$, as can be seen by comparing the fourth or fifth panel and the bottom panel. 

The changes in temperature at these small timescales are connected to the local particle heating processes at magnetic structures. The signatures in proton VDF (velocity distribution function) of local heating for these two representative sub-intervals are reported in Fig. \ref{fig:six}, where we show the reduced  VDF ($f(t,v_\parallel),f(t,v_{\bot_{1,2}})$ for the non-Alfv\'enic (left panel) and for the Alfvénic (right panels) sub-intervals described above (and reported in Fig.~\ref{fig:five}). We also show the reconstructed 3D VDFs in the plane $\{(v_\parallel,v_{\bot_{1}},v_{\bot_{2}})\}$, and its projection into the $(v_\parallel,v_{\bot_{1,2}})$ planes, at the bottom panels for each type of wind.  The reduced VDF has been obtained by first transforming the initial 3D energy distribution function from the $\{E,\theta,\varphi\}$ space (energy, elevation, and azimuth) to velocity space in the field-aligned coordinate system $\{v_\parallel,v_{\bot_1},v_{\bot_2}\}$.  Details can be found in Appendix \ref{appendix:graph}.
 
As can be seen from Fig.~\ref{fig:six}, the proton VDF shows strong deviations from local thermal equilibrium and gyrotropy in the vicinity of large PVI values. We observe that protons have enhanced perpendicular velocities in those regions, and parallel propagating proton beams around the local Alfv\'en speed are observed near discontinuities in both Alfvénic and non-Alfvénic intervals. We note some differences in those two distributions. In the non-Alfv\'enic case the beam is hot in the perpendicular direction. Such proton beams with large $T_{\bot,b}/T_{\bot,c}$ are consistent with the ``hammerhead'' distribution \citep{verniero2022strong}. In the Alfv\'enic case the beam is instead cooler and more focused in the parallel direction. This can be seen by inspecting Fig.~\ref{fig:six_new}, which shows a scatter plot of the beam-to-core drift velocity normalized by Alfv\'en speed versus $(T_{\bot,b}/T_{\bot,c})^*$, which is  the ratio of the perpendicular temperature of the beam and the core ($T_{\bot,{b}}/T_{\bot,{c}}$) normalized by the combined beam+core $T_{\bot,{b+c}}/T_{\parallel,{b+c}}$ values. Details on the fitting method can be found in the appendix \ref{appendix:fitting}. For reference, the crosses show data point at $t=$ 03:35:39~UT (non-Alfv\'enic wind) where $(T_{\bot,b}/T_{\bot,c})^*=2.81$ and at $t=$ 01:22:33~UT, where $(T_{\bot,b}/T_{\bot,c})^*=1.16$. To help intuition, we have split the scatter plot into four quadrants that we use as a reference to characterize the shape of the VDF. On the right side of the diagram, the values on the lower quadrant represent VDFs with a focused beam, and values on the upper quadrant represent VDFs with hotter beams or “hammerhead” distributions. The values on the left side, the beam and core are not well separated, and one can arbitrarily attribute them to distributions with no beams. A sketch of the shape of the distribution as one moves above and below the line $({T_{\bot_{b}}/T_{\bot_{c}}})^*=1$ is also presented on the bottom right corner of Fig.~\ref{fig:six_new}. Fig.~\ref{fig:six} suggests that beams in the Alfv\'enic and non Alfv\'enic wind have different origins and display distinct properties.  

\begin{figure}
  \centering
  \includegraphics[width=0.5\textwidth]{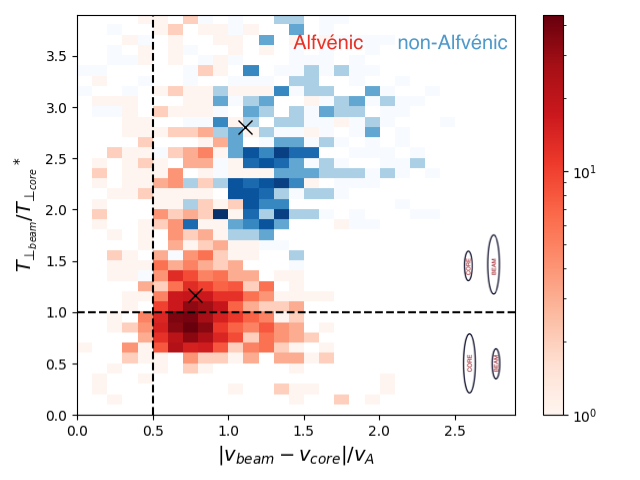}
  \caption{Scatter plot of the beam-to-core drift velocity normalized by Alfv\'en speed versus the ratio of the perpendicular temperature of the beam and the core for the same Alfvenic and non-Alfvenic intervals presented before.  The black crosses in the blue and red contours correspond to the local values at $t=$ 03:35:39~UT and $t=$ 01:22:33~UT, respectively.}
\label{fig:six_new}
\end{figure}

\section{Hybrid simulations}
\label{sect3}
\subsection{Initial conditions}
We performed hybrid-kinetic simulations to investigate proton heating at different types of magnetic structures under different plasma conditions, and compared our results with observations. We considered a quasi-neutral hybrid plasma model with massless and isothermal electrons, while protons are modeled as kinetic particles using the particle-in-cell method \citep{matthews1994,Franci_2018}. We adopted the following normalization: lengths are normalized to the proton inertial length $d_i =  c/\omega_{p}$ with $\omega_p = (4 \pi ne^2/m_i)^{1/2}$ the proton plasma frequency and the time is normalized to the inverse of the proton gyrofrequency $\omega_{ci}^{-1} = (eB_0/m_i c)^{-1}$ and velocities to the Alfv\'en speed $v_A = B_0/(4 \pi n_0 m_i)^{1/2}$ with $B_0$ the mean magnetic field. The plasma beta for both ions and electrons is defined as $\beta_{p,e}=8\pi n_0 T_{p,e}/B_0^2$. To avoid energy accumulation at the grid scale, we have included explicit resistivity with a corresponding dissipation length ($l_d$) related to the Reynolds number and the box size ($L$) through $R_e \sim (L/l_d)^{4/3}$, which is chosen to be greater than the grid size.

We have considered two different setups in 2.5D geometry  characterized by two different types of turbulent fluctuations resembling non-Alfv\'enic and Alfv\'enic wind, namely, $(i)$ a decaying 2D balanced turbulence simulation, and $(ii)$ a wave simulation with an initial parallel propagating, Alfv\'enic broadband spectrum. For case $(i)$, hereafter referred to as ``turbulence simulation'', we considered an initial large-amplitude fluctuation with $\sigma_c \approx 0$, in energy equipartition and with an out-of-plane guide field $\mathbf{B_0} = B_0 \mathbf{\hat{z}}$. The initial condition consists of large-amplitude perturbations of perpendicular wave modes in the range $ 0.196<k_\bot d_i< 0.49$ with random phases, similar to previous work \citep{servidio2012local,franci2015high,cerri2017reconnection}. For case $(ii)$, referred to as ``wave-like simulation'', we considered an in-plane mean-field $\mathbf{B_0} = B_0 \mathbf{\hat{x}}$ and the initial perturbation satisfies the Wal\`en relation in the dispersionless limit $\delta {\bf u} = -(\omega_0/k_0) \delta {\bf b}$. The wave frequency $\omega_0$ is determined from the normalized dispersion relation $k_0^2=\omega_0^2/(1-\omega_0)$ for left-handed polarized parallel propagation waves. This initial condition corresponds to a broadband Alfv\'enic fluctuation comprised of parallel modes in the range $ 0.196<k_\parallel d_i< 0.49$ (see also  \citet{Malara2000,matteini2010kinetics,Gonzalez_2021}). In both cases, an initially isotropic and Maxwellian plasma with proton beta $\beta_i=0.5$ and equal ion and electron temperature $T_i/T_e=1$ are considered. The  guide  field is $B_0=1$ and the same root-mean-square (rms) of the magnetic fluctuations $\delta b_{rms}/B_0 = 0.63$ is chosen for both simulations. We adopt periodic boundary conditions and fix a box of side $L=128 \ d_i$ by using $1024^2$ grid points with mesh size $0.125 \ d_i$ and $8000$ particles per cell. 

Because of the different geometry adopted, interactions between fields and particles are fundamentally different in the two setups \citep{gary2020particle}. Naturally, our simulations are not meant to represent realistically the two types of wind (both requiring 3D fields), but rather to isolate different processes that  might be dominant in each type of wind and that bear specific signatures in phase space. In section \ref{overview_numerical} we describe the evolution of fields and particles in the two setups. In section \ref{PVI} we discuss the correlation between anisotropy and PVI, and the signatures in phase space of proton heating/acceleration in correspondence with at large PVI values.

\subsection{Overview of numerical results}
\label{overview_numerical}

\begin{figure}[t!]
 \centering
 \includegraphics[width=\textwidth]{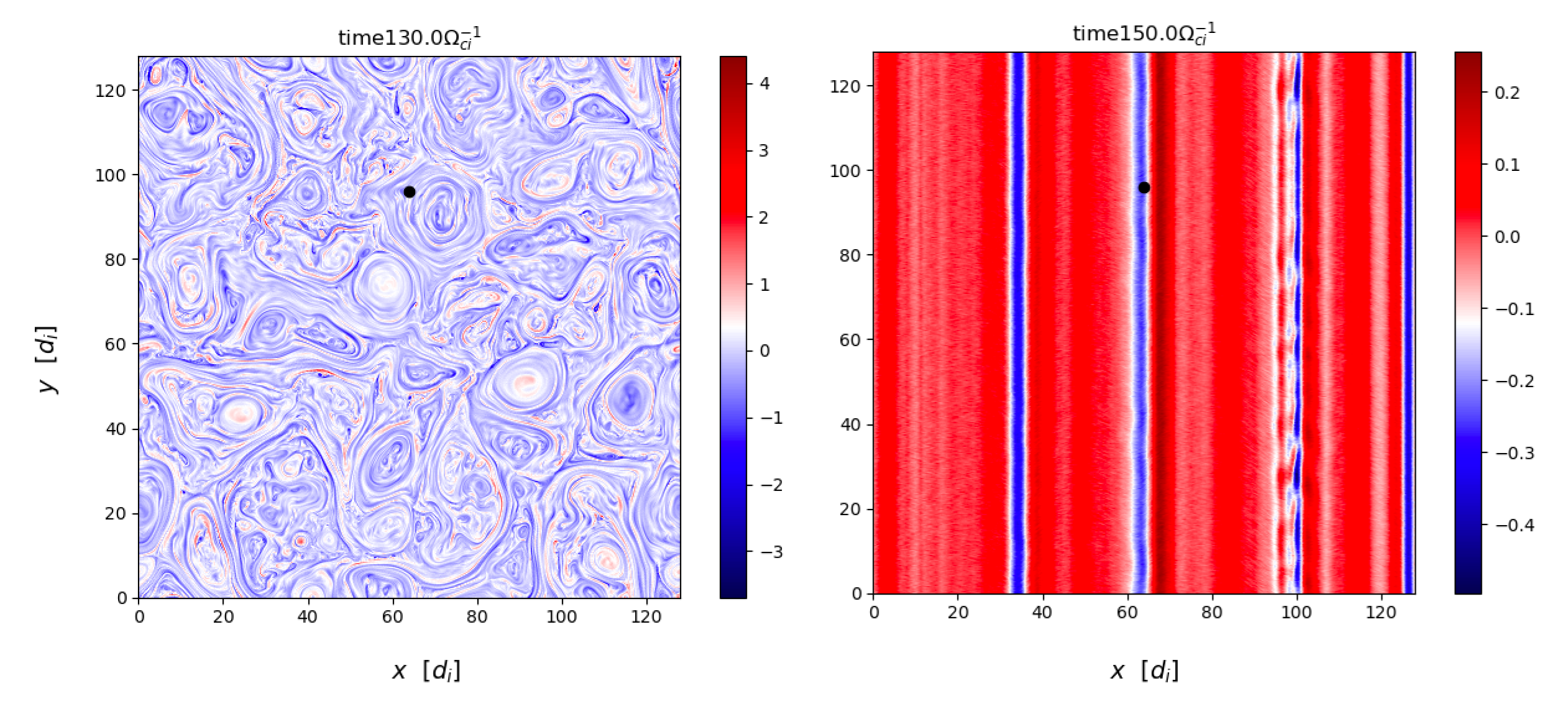}
 \caption{Contour plot of the out-of-plane current density $J_z$ for the turbulence simulation (left) and the wave-like simulation (right). The black dot shows the location at which the time series shown in Fig.~\ref{fig:eight} have been taken.}
\label{fig:Fig_space_simulations}
\end{figure}

In the turbulence simulation, once the system reaches the fully developed turbulent state, the energy stored in the fields is progressively converted into thermal energy, resulting in an average (over the spatial domain) preferential perpendicular heating with $\langle T_\bot/T_\parallel \rangle_{box}=1.37$ at $t=120 \omega_{ci}^{-1}$.  The wave-like simulation, by contrast, displays a strong enhancement of the spatially averaged $T_\parallel$, and the temperature anisotropy reaches $\langle T_\bot/T_\parallel\rangle_{box}=0.6$ at $t=180 \omega_{ci}^{-1}$\footnote{Additional plots and movies showing the global dynamics and overall proton heating in the simulations can be found in the supplemental material.}.

These different heating processes in the two setups are expected,  since the turbulence simulation leads to a well-developed magnetic energy spectrum perpendicular (in $k$-space) to the guide field, allowing for more channels of particle heating, such as stochastic heating and particle scattering at current sheets \citep{cerri2021stochastic,sioulas2022particle}. Those processes are suppressed in the wave-like simulation due to the different geometry adopted. On the other hand, in the wave-like simulation, the available energy carried by the wave is converted into kinetic and thermal energy via the rapid disruption of the wave packet mediated by wave steepening along the guide field (i.e., through the formation of gradients parallel to ${\bf B}_0$); this leads to both the formation of a field-aligned proton beam at about the local Alfv\'en speed, and local perpendicular heating at the steepened edges \citep{Gonzalez_2021}. This mechanism is in turn suppressed in the turbulence simulation. Figure~\ref{fig:Fig_space_simulations} shows the contour plot of $J_z$ for the two setups to show the different types of structures that form nonlinearly.

\begin{figure}[t!]
 \centering
 \includegraphics[width=\textwidth]{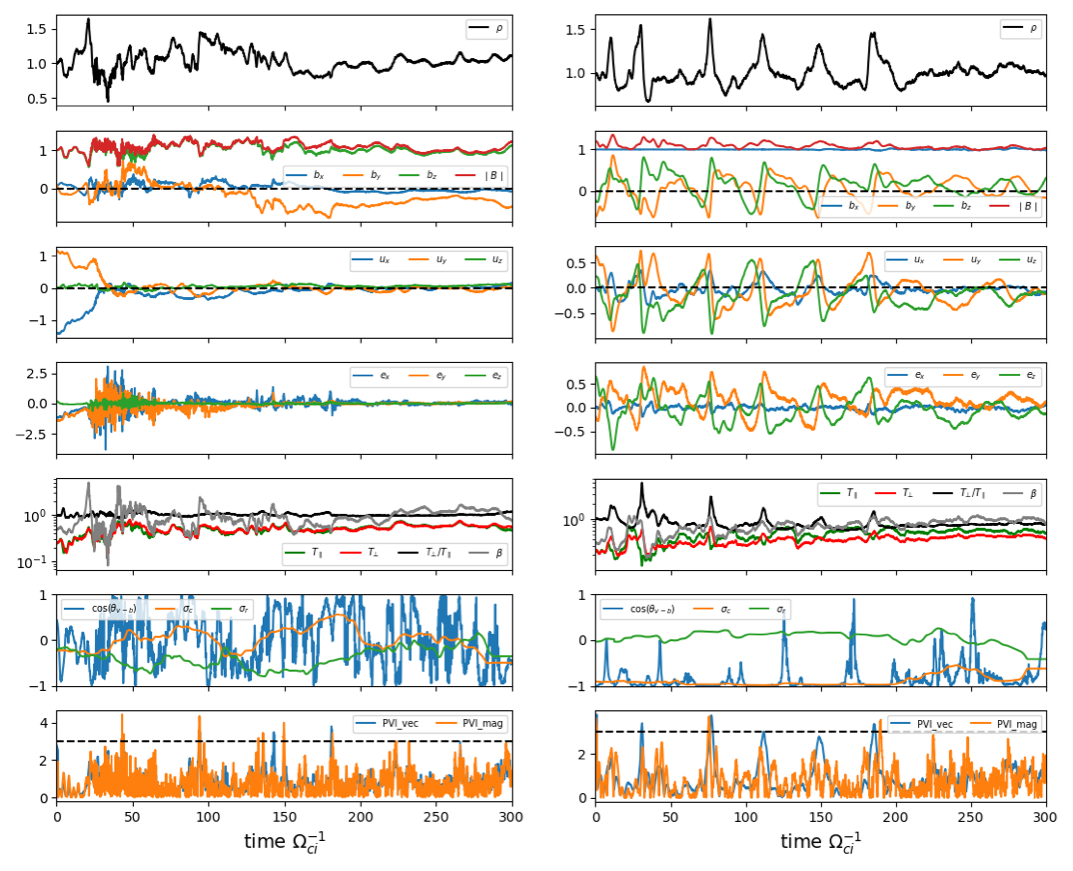}
 \caption{Single-point time evolution of fields from the turbulence simulation (left panels) and wave-like simulation (right panels). From top to bottom: proton density, magnetic field, proton bulk velocity, electric field,  parallel and perpendicular temperature, temperature anisotropy, and plasma beta. Finally, the last two panels show $\sigma_c$, $\sigma_r$ and $\cos\theta_{vb}$, and the PVI values.}
\label{fig:eight}
\end{figure} 

To make a comparison with PSP data, in Figure~\ref{fig:eight} we show the time series of fields and plasma quantities in the two setups. The time series have been taken at the location marked by the black dot in Figure~\ref{fig:Fig_space_simulations}. From top to bottom, Figure~\ref{fig:eight} shows the time series of $n$, ${\bf B}$ and $|B|$, ${\bf u}$, the electric field ${\bf e}$, $T_\bot$, $T_\parallel$ and $\beta$, $\cos \theta_{vb} $, $\sigma_r$ and $\sigma_c$, and the PVI values (see section \ref{PVI} for details on the PVI analysis). The turbulence simulation (left panels), is characterized, on average by $|\sigma_c|<0.5$ and  $\sigma_r<0$ (predominantly). The wave-like simulation (right panel) has $\sigma_r\approx0$ and $|\sigma_c|\approx1$. As can be seen,  temperature enhancements occur in correspondence with large PVI values in both cases. In the turbulence case, those large PVI values are associated with structures such as flux tubes, current sheets and small-scale plasmoids. These structures have a strong perpendicular electric field, and some also display large out-of-plane electric and magnetic fields (along the \textit{z}-axis) with strong compressibility, mediating particle acceleration and heating \citep{dmitruk2004test,Wan2015,comisso2022ion}.  As discussed above, in the wave-like simulation large PVI values are associated with the steepened edges of Alfv\'enic fluctuations. Also, rapid and large variations of $\cos\theta_{ub}$ near these structures can be identified (sixth panel), similar to what is observed during the Alfvénic sub-interval presented in section \ref{proton_kinetic_PSP}. In contrast to the magnetic structures in the turbulent simulation, the steepened edges are mostly rotational discontinuities with an embedded compressive component \citep{matteini2010kinetics, Gonzalez_2021}. 

\subsection{Correlations between magnetic structures and proton kinetic features}
\label{PVI}

The PDFs conditioned on $PVI$ and $PVI$-$MAG$ are shown in Fig.~\ref{fig:seven} in the same format as Fig.~\ref{fig:three}. To obtain the PDFs, we used a similar methodology to that employed for PSP data, but we have taken $\tau= 1 \omega_{ci}^{-1}$ since we have enough resolution to inspect proton scales. We have then taken the time series of plasma quantities at a resolution of $0.1 \omega_{ci}^{-1}$ at $10^4$ fixed equidistant points in the simulation domain.
\begin{figure}
  \centering
 
\includegraphics[width=\textwidth]{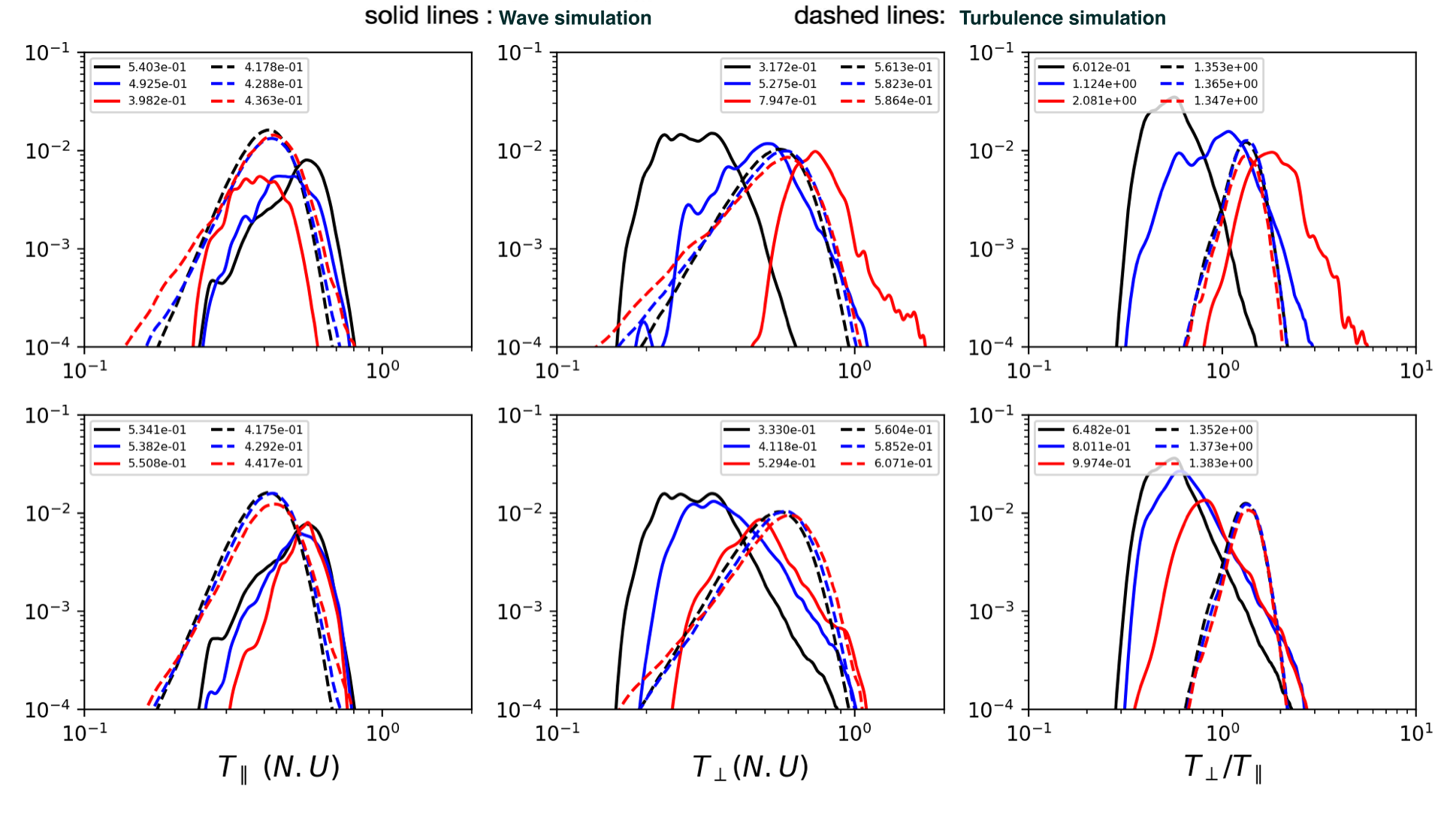}
  \caption{Conditional PDF with respect to $PVI$ (top panels) and $MAG$-$PVI$ (bottom panels) of proton temperatures using simulation data: PDF of parallel proton temperature $T_\parallel$ (in normalized units; left panels); PDF of perpendicular proton temperature $T_\bot$ (middle panels); PDF of proton temperature anisotropy $T_\bot/T_\parallel$ (right panels).  The solid lines correspond to the wave-like simulation and the dashed lines to the turbulence simulation. Different color lines show  PDF that corresponds to different ranges of $PVI$ and $MAG$-$PVI$: Black lines ($0<PVI\le1.5$), blue lines ($1.5<PVI\le3.0$), and red lines ($PVI>3.0$). The numbers on the legends are the mean value of each distribution.}
\label{fig:seven}
\end{figure}

In general, there is a positive correlation between PVI values with $T$ and $\delta T$  (not shown), indicating that the hottest protons are found locally near small-scale structures (in this case represented by $PVI>3 $ values). Although the turbulence simulation is hotter than the wave-like simulation (this is true both locally at discontinuities and globally over the simulation domain), protons remain close to Maxwellian in the turbulence simulation. On the contrary, in the wave-like simulation, particles develop large anisotropies localized at small-scale structures. In that case such a local anisotropy, reaching an average value of  $T_\bot/T_\parallel\approx2$ for $PVI>3$, is caused by strong particle pitch-angle scattering at the steepened fronts, resulting in effective local proton perpendicular energization \citep{Gonzalez_2021,malara2021charged,gonzalez2023}. Since this effect is localized at the steepened edges, the background temperature (represented by the lower PVI range) and the spatially averaged temperatures over the entire domain display an opposite anisotropy, $T_\bot/T_\parallel<1$, as discussed previously, due to the formation of a field-aligned beam that fills the entire simulation domain.

\begin{figure}[b!]
  \centering
  \includegraphics[width=\textwidth]{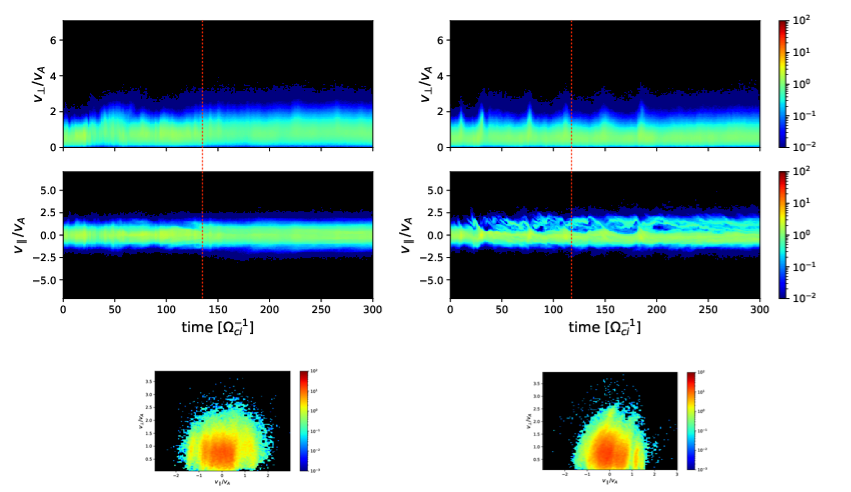}
  \caption{Single-point proton VDF from simulations. The reduced proton VDF of perpendicular ($f(t,v_{\bot_1,\bot_2},t)$) and parallel $f(t,v_\parallel)$ of proton velocity for the turbulence simulation (left panels) and wave-like simulation (right panels). The bottom panels show the reduced VDF $f (v_\parallel,v_\bot)$ at the time indicated by the red dashed line.}
\label{fig:nine}
\end{figure}

Some differences between  $PVI$ (Fig.~\ref{fig:seven}, top panels) and  $MAG$-$PVI$ (Fig.~\ref{fig:seven}, bottom panels) results can be noticed. In the wave-like simulation, the magnetic structures correspond essentially to rotational discontinuities with relatively small variations in the magnitude of $B$ and, thus, the correlation with the various PVI ranges is reduced in the $MAG$-$PVI$ case. This can be seen from the relative temperature variations between the smallest and the largest range of PVIs, showing the aforementioned trend ($PVI$:  $\Delta T_\parallel/T_\parallel\simeq 26 \%$ and $\Delta T_\bot/T_\bot\simeq 150\%$; $MAG$-$PVI$: $\Delta T_\parallel/T_\parallel\simeq 3 \%$ and $\Delta T_\bot/T_\bot\simeq 59\%$). For the turbulence simulation, on the contrary, higher relative temperature increase is found in the $MAG-PVI$ case, similarly to observations (PVI: $\Delta T_\parallel/T_\parallel\simeq 4.44\%$ and  $\Delta T_\bot/T_\bot\simeq 4.47\%$; $MAG$-$PVI$: $\Delta T_\parallel/T_\parallel\simeq 5.79\%$ and $\Delta T_\bot/T_\bot\simeq 8.33\%$). 

In Fig.~\ref{fig:nine}, top two panels, we show the gyroaveraged reduced perpendicular and parallel VDF ($f(t,v_{\bot})$ and $f(t,v_{\parallel})$, respectively and, in the bottom panels,  we show the reduced VDF $f(v_{\parallel},v_{\bot})$ at the time indicated by the dashed line, marking the occurrence of an event with $PVI>3$ in each case. The left column corresponds to the turbulence simulation and the right to the wave-like. As can be seen, magnetic discontinuities/intermittent structures are characterized by non-Maxwellian features, namely, temperature anisotropies and beams at the Alfv\'en speed. However, the turbulence simulation undergoes perpendicular heating in both the core and the beams, so that $T_\perp\gtrsim T_\parallel$ locally and on average, and beams are ``hot".  The wave-like simulation instead displays a strong perpendicular heating of the core, localized at the discontinuities, and a cold beam at the Alfv\'en speed (more focused in the perpendicular direction than in the turbulence setup).  

To summarize, our simulations show that, in analogy with PSP data, higher temperatures and temperature variations are found at higher PVI values in both setups (both winds), and that colder beams (as found in Alfv\'enic wind) are associated with the steepening of Alfv\'enic fluctuations while hot beams (as found in non-Alfv\'enic wind) are generated in low-cross helicity turbulence. The development of a standard turbulent cascade also favors preferential perpendicular heating (as in the non-Alfv\'enic wind). Even though our simulations highlight  the different origins and properties of proton beams in the two types of setups/winds, and qualitatively reproduce observed properties of non-Alfv\'enic wind, the do not reflect all the local trends of temperatures at magnetic structures observed in the Alfv\'enic wind. In particular, the PVI-temperature analysis in the wave-like simulation shows 
a much larger relative increase of perpendicular temperature than observed (larger than the increase in parallel temperature), and compressibility of the Alfv\'enic wind at small sacels is also not well reproduced numerically. These discrepancies are likely due to the geometry contraints. However, we also mention the possibility that instrument resolution may also play a role in underestimating strong temperature variations at small scales.   

Before concluding we also remark that our simulations do not reproduce differences between the two types of winds in their average properties, such as the fact that the Alfv\'enic wind is hotter, and that the global (average) anisotropy is $T_\perp/T_\parallel<1$. This is expected since our setups cannot reproduce fully 3D dynamics. Furthermore, such differences might well be related to different coronal origins and/or a different radial evolution of Alfv\'enic and non-Alfv\'enic wind, an aspect that is not addressed in this work.

\section{Summary and conclusions}
\label{sect4}

We have studied the correlation between proton temperatures and magnetic discontinuities/ intermittent structures in different solar wind turbulence conditions (high and low cross helicity, i.e., Alfv\'enic and non-Alfv\'enic wind) by using PSP observations from E6-E9 and hybrid-kinetic simulations. Our main results and conclusions are summarized as follows:

\begin{itemize}
    \item At large (MHD) scales, the Alfv\'enic wind is much less compressible than the non-Alfv\'enic wind (see Fig.~\ref{fig:two}). However, our $PVI$ and $MAG$-$PVI$ analyses show that the hottest protons are localized at kinetic-scale compressible structures in both types of wind (see Fig.~\ref{fig:three}). 
    \item There is a statistical correlation between the highest proton total temperature and coherent structures (quantified by PVI values), consistent with previous studies \citep{Qudsi_2020,sioulas2022}. Furthermore, the Alfv\'enic wind shows a preferential enhancement of $T_\parallel$ as smaller scale structures are considered, whereas the non-Alfv\'enic wind shows a preferential enhancement of $T_\bot$ (see Fig.~\ref{fig:three}). 
    \item Proton beams are ubiquitous in both types of wind (leading to an average anisotropy $T_\parallel > T_\bot$.) However, local kinetic features of proton VDFs differ in the two winds (see Fig.~\ref{fig:six} and \ref{fig:six_new}). The non-Alfv\'enic wind is characterized by ``hot" beams ($T_{\bot,b}/T_{\bot,c}^*\gtrsim2$) resembling the ``hammerhead" distributions. The Alfv\'enic wind is characterized by ``colder" beams ($T_{\bot,b}/T_{\bot,c}^*\lesssim 1$).

   \item We find some similarities between the hybrid-kinetic simulations and in-situ measurements despite the limitations of the reduced geometry adopted. The field aligned proton beams that develop in our simulations display distinct features, supporting the idea that proton beams in Alfv\'enic and non-Alfv\'enic wind have different properties and different origins. Simulations suggest that the development of a perpendicular cascade, favored in balanced turbulence, allows a preferential relative enhancement of $T_\bot$, and the formation of hot beams via nonlinear dynamics and reconnection. On the contrary, cold field-aligned beams are favored by Alfv\'en wave steepening (see Fig.~\ref{fig:seven} and Fig.~\ref{fig:nine}). 
\end{itemize}

Additionally, we have shown for the first time 3D proton VDFs from PSP displaying non-Maxwellian and non-gyrotropic features near discontinuities, warranting a more general approach to fit VDFs than the widely adopted fit to bi-Maxwellians. Furthermore, non-Maxwellian and non-gyrotropic  proton VDFs around discontinuities/intermittent structures are found in both winds, confirming that nonlinearities and strong deviations from non-thermal distributions are intrinsically related in collisonless plasmas \citep{valentini2014hybrid}, resulting in a  universal heating channel regardless of the Alfvénic properties of the solar wind.
In conclusion, this work contributes to understanding the distinctive role of coherent structures in heating collisionless plasmas. To gain a comprehensive understanding, our simulation results should be extended to include three-dimensional effects and encompass a broader range of initial Alfvénic properties rather than the extreme cases considered here ($\sigma_c \simeq 0$ and $\sigma_c \simeq -1$) This will be the subject of future investigations.

\section*{Acknowledgements}
\acknowledgements{This research was supported by NASA grant \#80NSS\-C18K1211 and NSF CAREER award 2141564.  We acknowledge the Parker Solar Probe (PSP) mission for the use of the data publicly available at \href{https://spdf.gsfc.nasa.gov/}{the NASA Space Physics Data Facility} and the Texas Advanced Computing Center (TACC) at The University of Texas at Austin for providing HPC resources that have contributed to the research results reported within this paper. Simulations have been run on the Frontera supercomputer  http://www.tacc.utexas.edu. JLV acknowledges support from NASA PSP-GI 80NSSC23K0208 and NASA LWS 80NSSC22K1014.}

\bibliography{addon}

\appendix
\renewcommand{\thefigure}{A\arabic{figure}}
\setcounter{figure}{0}

\section{VDF interpolation method}
\label{appendix:graph}
To visualize the proton velocity distribution in velocity space we mapped the initial 3D energy distribution function from energy, elevation, and azimuthal angle coordinates $(E=1/2mv^2,\theta,\phi)$ into velocity-space coordinates in the instrument frame ($v_x,v_y,v_z$) using transformation from spherical to cartesian coordinates:
\begin{equation*}
v_x = v \cos\theta \cos\phi, \ \ \ \ \ v_y = v \cos\theta \sin\phi, \ \ \ \ \ v_z = v\sin \theta, 
\end{equation*}

with $v$ the amplitude of the velocity field. The $v_z$ component is defined as the rotational symmetry axis of the instrument, $v_x$ direction points toward the sun, and the $v_y$ component is then defined orthogonal to them. However, the results presented in section~\ref{sect2} are shown in the field-aligned frame, which was obtained by applying a rotation matrix to the proton velocity vectors in the instrument frame: 

\begin{equation*}
\begin{bmatrix}
v_\parallel\\v_{\bot_1}\\v_{\bot_2}
\end{bmatrix}
 =
\begin{bmatrix}
\cos \psi & -k_z \sin \psi & k_y \sin \psi \\
k_z \sin \psi & k_y^2 + k_z^2 \cos \psi & k_y k_z (1 - \cos \psi) \\
-k_y \sin \psi & k_y k_z (1-\cos \psi) & k_z^2 + k_y^2 \cos \psi
\end{bmatrix}
\begin{bmatrix}
    v_x\\v_y\\v_z
\end{bmatrix},
\end{equation*}

The above expression is obtained by using the Euler-Rodriguez formula that corresponds to a rotation by an angle $\psi$ about an axis defined by the unit vector $\mathbf{\hat{k}}$. Here we chose this axis to be defined as $\mathbf{\hat{k}} = [0, -\hat{b_z}, \hat{b_y}]$ and $\psi$ the angle between the instrument frame pointing to the sun and the magnetic field vector $\mathbf{\hat b}$. Further details can be found in \citet{woodham2021enhanced}.
To obtain a reduced representation (1D/2D) of the integrated velocity distribution function that accurately accounts for the different velocity ranges at different planes, we used 3D linear interpolation on a mesh with a fixed velocity range. This method ensures the proper integration along the preferred component that appropriately adds up the number of counts in each velocity bin. The interpolation method is commonly used to approximate a function from a set of discrete data and it returns a callable function that can be used to evaluate the interpolated function at any point within a defined interval. We employed the cubic interpolation method from the \texttt{griddata} function in the \texttt{scipy.interpolate} module \citep{2020SciPy-NMeth}. For illustration, Figure~\ref{appendplot} presents a 3D render of the original and the interpolated proton VDF. The Python code can be found in the following~\href{https://github.com/caangonzalez1/data_analysis/blob/main/SPANi_3DVDF.ipynb}{repository}.

\begin{figure}[htp]
  \centering
  \includegraphics[width=\textwidth]{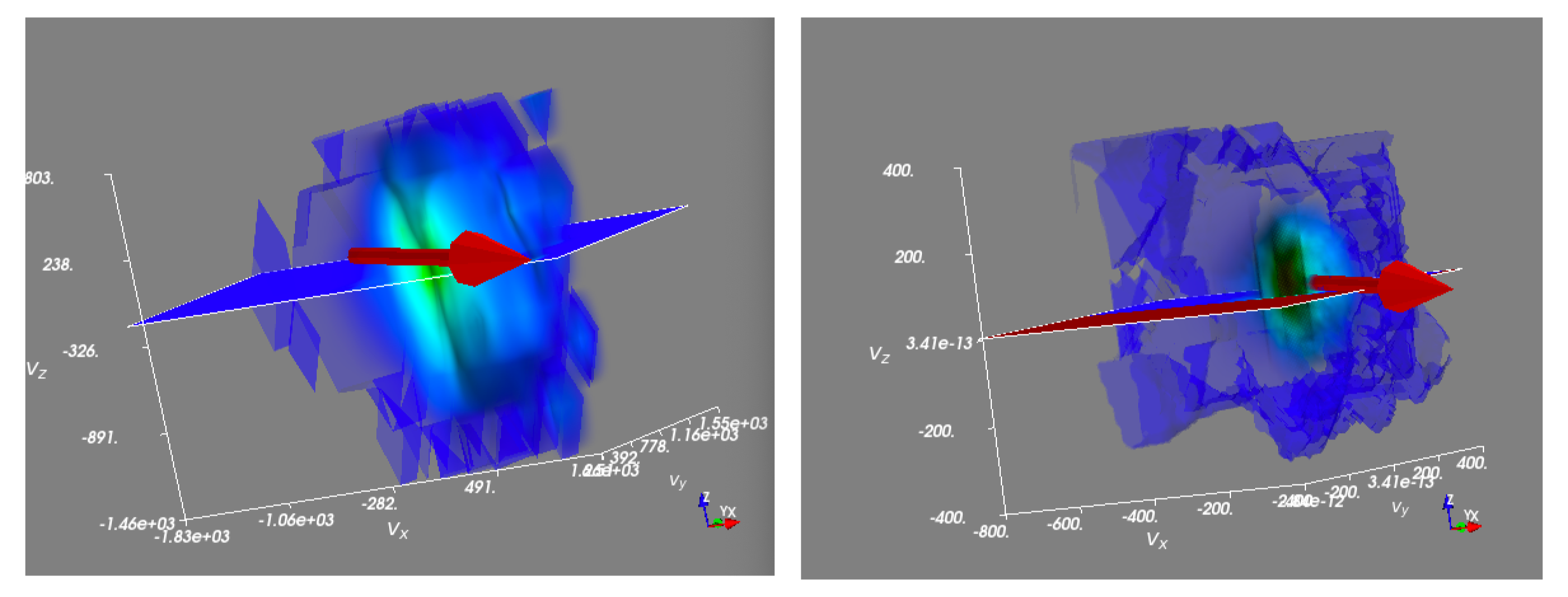}
  \caption{Illustration of the interpolation method. Several VDFs that correspond to different measurements in azimuthal angles (represented by the range of colors from yellow to purple) and for different elevation angles (arranged from left to right). For each VDF,  it is shown the original signal (solid lines) and the interpolated signal (dashed lines).}
\label{appendplot}
\end{figure}

\section{Fitting method}
\label{appendix:fitting}
To obtain additional information on the proton VDFs, the SPAN-I L2 sf00 data product was fit to a sum of two 3D bi-Maxwellians, one for the ``core" and one for ``beam," following a similar procedure as described in \citep{Verniero_2020}. Note that the ``beam" was constrained to lie parallel to the mean magnetic field and the  ''core" was labelled as the peak in phase-space density. Since SPAN-I is partially obstructed by PSP's Thermal Protection Shield, only partial moments of the VDF are obtained. We therefore only include fits that are at least 80\% in the field-of-view (FOV). We quantify this by first computing moments of the distribution from a VDF reconstructed from the fitting parameters. Next, we take the sum of the fitted beam and core densities and divide by the reconstructed moment density. To exceed a FOV threshold of 80\%, this number must exceed 0.8.

\end{document}